\documentstyle[12pt]{article}
\def\be{\begin{equation}}
\def\ee{\end{equation}}
\def\bea{\begin{eqnarray}}
\def\eea{\end{eqnarray}}

\begin{document}

\begin{center}
{\Large{\bf The Nambu-Goto and Polyakov 
Strings via the Relativistic Particles}}

\vskip .5cm
{\large Davoud Kamani}
\vskip .1cm
 {\it Faculty of Physics, Amirkabir University of Technology 
(Tehran Polytechnic)
\\  P.O.Box: 15875-4413, Tehran, Iran}\\
{\sl e-mail: kamani@aut.ac.ir}
\\
\end{center}

\begin{abstract}

We assume the bosonic string is a composite object
of the relativistic particles. The behavior of the
relativistic particles in a curve enables us to obtain the
Nambu-Goto and the Polyakov actions of the bosonic string.
We observe that the particles of these strings move with
non-constant speeds along them. 

\end{abstract}



\vskip .5cm
\newpage

\section{Introduction}

There have been some attempts to understand the string bits 
\cite{1}-\cite{13}.
The origin of the idea of the string bits can be traced to the earliest days
of the dual models \cite{14,15}. 
Recently, the string bits have been discussed 
in the pp-wave background \cite{1,2,3,4,5}. The idea of the
string bits also appeared in the 'tHooft's works \cite{6}. 
There are string bit
models which are based on supersymmetric quantum mechanics \cite{7}. 
Some other
pictures for strings in terms of the bits language, corresponding to
the quantum gravity, have been studied \cite{8,9,10}.
For more investigation on the string bits also 
see the Ref. \cite{16,17,18,19}.

Classically, the Nambu-Goto and Polyakov strings are equivalent.
We are motivated to find difference of them, classically, in terms
of their bits. That is, classical behavior of a bit of the Nambu-Goto
string is different from the behavior of a bit of the Polyakov string.
We consider relativistic particles as the string bits.

We shall give a picture for the bosonic strings in terms of
the relativistic point particles. Although the actions of a string are
analogue of the actions of a relativistic point particle, however,
apparently there is no relation between these particles and
strings. We connect the actions of a 
particle to the Nambu-Goto and Polyakov
actions of a string. This connection 
enables us to study the motion of a string bit
along the string. We observe that the string particles are not at
rest relative to the string. The motion of a bit of the Nambu-Goto
string is different from the motion of a bit of the Polyakov string.
As physical results, these motions determine the string
mass density and, put an upper bound on the string 
length scale. Some quantum considerations 
simplify the classical speed formulas, and reveal non-uniform 
distribution of the particles on the string.

In all string-bit models when the number of bits goes to infinity the
usual string is recovered. Similarly, in our model, string is
a continuum of infinite relativistic particles.

This paper is organized as follows. In section 2, the action of a
moving particle in the string will be studied. In section 3, the
actions of the Nambu-Goto string and the Polyakov string through
their particles will be obtained. In section 4, the motion of the
particles of the Nambu-Goto string and the Polyakov string,
relative to these strings, will be analyzed. 
In section 5, by canonical quantization, the particle speeds will be
simplified.
\section{Action of a relativistic point particle in the string}

\subsection{Relativistic point particle}

Consider a relativistic point particle in the $D$-dimensional
curved spacetime ($e.g.$,
see \cite{20}). Let the coordinates of this spacetime
be $\{X^\mu | \mu=0,1,...,D-1\}$. The curve $X^\mu(t)$, with the
parameter ``$t$'' along it, describes the world-line of the
particle. The Nambu-Goto form of the action of the particle is given by
\bea
S^{(pp)}_{NG} = -m \int dt \bigg{(} -G_{\mu\nu}(X)
\frac{dX^\mu}{dt} \frac{dX^\nu}{dt} \bigg{)}^{1/2} ,
\eea
where $G_{\mu\nu}(X)$ is the metric of the curved spacetime
with one negative eigenvalue, corresponding to the time direction, 
and $D-1$ positive eigenvalues, associated to the space directions,
and $m$ is the particle mass. In fact, this action shows the invariant
proper time along the particle world-line.

Introducing an independent degree of freedom, $i.e.$
world-line metric $h(t)$, leads to the Polyakov form of the particle action
\bea
S^{(pp)}_P = \frac{1}{2} \int dt \bigg{(} h^{-1}(t)G_{\mu\nu}(X)
\frac{dX^\mu}{dt} \frac{dX^\nu}{dt} - h(t)m^2\bigg{)}.
\eea
The action (2) works for both massive and massless particles, while
the action (1) only is appropriate for the massive particles.
Both of them have world-line reparametrization invariance. Note that under
the reparametrization $t \longrightarrow t'(t)$ the world-line
metric $h(t)$ transforms to $h'(t')$ such that
\bea
h'(t')dt' = h(t)dt .
\eea

Each of the actions (1) and (2) has its own advantages.
However, at the classical level they are equivalent. That is,
use the equation of motion of $h(t)$,
\bea
h^2(t)=-\frac{1}{m^2}G_{\mu\nu}(X)
\frac{dX^\mu}{dt}\frac{dX^\nu}{dt},
\eea
to eliminate $h(t)$ from (2), the
action $S^{(pp)}_P$ reduces to the action $S^{(pp)}_{NG}$. From now on we
call the particles with the actions (1) and (2) as the Nambu-Goto
particle and the Polyakov particle, respectively.
\subsection{Relativistic point particle in the string}

The string shape is a mathematical curve $C$. 
We assume the string is a composite object of the relativistic
particles, distributed along the curve $C$.
Now we obtain the action of a point particle which moves along
the curve $C$. Each particle on 
the string curve is distinguished by its position,
$i.e.$ the worldsheet coordinate $\sigma$. All particles on the
string curve have the common parameter $\tau$ on their world-lines, where
$\tau$ is the time coordinate of the string worldsheet. This enables us
to choose the parameter $\tau$ on the particle world-line.
The various world-lines are distinguished by various values of $\sigma$.
Thus, the string worldsheet is constructed by infinite number of the 
particle world-lines.
In this picture, the particle ``$p$'' has a worldsheet-line in the
string worldsheet, $i.e.$ $ \sigma_p (\tau)$, and a world-line in
the spacetime, $i.e.$ $X^\mu(\sigma_p(\tau) , \tau)$, where the
coordinates $X^\mu (\sigma , \tau)$ describe the string
worldsheet in the spacetime.

For an arbitrary world-line action of particle is given by
(1) or (2). For a particle in the string we should replace the variable
$\frac{dX^\mu}{dt}$ by $\frac{dX^\mu}{d\tau}$,
\bea
&~& \frac{d}{d\tau} X^\mu(\sigma_p(\tau) , \tau) = \partial_\tau X^\mu
+ {\dot \sigma}_p\partial_\sigma X^\mu,
\nonumber\\
&~& \partial_\tau X^\mu \equiv \frac{\partial}{\partial \tau'}
X^\mu(\sigma_p(\tau) , \tau')|_{\tau'=\tau} ,
\nonumber\\
&~& \partial_\sigma X^\mu \equiv \frac{\partial}{\partial
\sigma_p(\tau)} X^\mu(\sigma_p(\tau) , \tau) ,
\nonumber\\
&~& {\dot \sigma}_p \equiv \frac{d\sigma_p(\tau)}{d\tau} .
\eea
Thus, the motion of the particle in the spacetime 
is described by the motion of the
string which comes from the term $\partial_\tau X^\mu$, and the
motion of the particle relative to the string which is given by
the term ${\dot \sigma}_p\partial_\sigma X^\mu$.

According to these, the actions of a particle
(moving in the string) are as in the following
\bea
S'^{(pp)}_{NG} = -m \int d\tau \bigg{(} -H^{ab}G_{\mu\nu}(X) \partial_a X^\mu
\partial_b X^\nu \bigg{)}^{1/2} ,
\eea
for the Nambu-Goto particle, and
\bea
S'^{(pp)}_P = \frac{1}{2} \int d\tau \bigg{(} h^{-1}(\tau)H^{ab}G_{\mu\nu}(X)
\partial_a X^\mu \partial_b X^\nu - h(\tau)m^2\bigg{)} ,
\eea
for the Polyakov particle. 
The indices ``$a$'' and ``$b$'' run over $\tau$ and
$\sigma$, and the symmetric matrix $H^{ab}$ has the definition
\bea
H^{ab}=\left( \begin{array}{cc}
1 & {\dot \sigma}_p \\
{\dot \sigma}_p & {\dot \sigma}^2_p
\end{array} \right).
\eea

The actions (6) and (7) under the world-line reparametrization
$\tau \longrightarrow \tau'(\tau)$ with $h'(\tau')d\tau' =
h(\tau)d\tau$, are invariant. Furthermore, by the equation of
motion of $h(\tau)$, $i.e.$, 
\bea h^2(\tau) =
-\frac{1}{m^2}H^{ab}G_{\mu\nu}(X) \partial_a X^\mu
\partial_b X^\nu ,
\eea
these actions classically are equivalent.
\section{String actions from the relativistic particles}

Apart from the square root,
the form of the action (6) is far from the Nambu-Goto action of
string. Similarly, although for the massless particle the action (7) has
the form similar to the Polyakov action of string, however, we cannot
identify $h^{-1}(\tau)H^{ab}$ with the intrinsic metric of the
string worldsheet, $i.e.$ $g^{ab}$. This is due to the fact that
$\det H^{ab}=0$, and hence $H^{ab}$ is not invertible, while
$g^{ab}$ is an invertible matrix with the inverse $g_{ab}$.

In fact, as we shall do, it is possible to obtain 
the Nambu-Goto string and the Polyakov
string from the actions (6) and (7), respectively.

\subsection{The Nambu-Goto string}

Consider particles in an interval
$(\sigma-\frac{d\sigma}{2} , \sigma+\frac{d\sigma}{2})$ of the string, which
is located at the position $\sigma$. Assume $d\sigma$ contains a large
number of the particles. This enables us to define a mass density for 
the string. Let $\rho$ be the mass density of the string,
$i.e.$ $\rho= dm/d\sigma$. 
Therefore, from the action (6) (after integration over $\sigma$) we obtain
\bea
S_{NG} = -\int d \sigma \int d \tau \rho
(-H^{ab}h_{ab})^{1/2},
\eea
where $h_{ab}$ is the pull-back of the spacetime metric $G_{\mu\nu}(X)$
on the string worldsheet
\bea
h_{ab} = G_{\mu\nu}(X) \partial_a
X^\mu \partial_b X^\nu .
\eea

The equation (10) describes the action of infinite number of the 
relativistic particles in a
curve. This system may be any one-dimensional physical object.
Demanding it be a special object imposes special behavior on the particles.
We want it be a Nambu-Goto string. This leads to the following behavior 
for each string bit which is given by the equation
\bea
H^{ab}h_{ab} =
\frac{1}{(2\pi \alpha'\rho)^2} \det h_{ab} .
\eea
This implies the Nambu-Goto action for the string
\bea
S_{NG} = -\frac{1}{2\pi
\alpha'} \int d \sigma \int d \tau \sqrt{-\det h_{ab}} .
\eea

In fact, if the relativistic particles along the string curve $C$
form the Nambu-Goto string, then the matrix $H^{ab}$ and the 
induced metric $h_{ab}$ have to relate to each other
through the equation (12). 
In other words, the equation (12) means that when a relativistic
particle is inside the Nambu-Goto string curve $C$, it should obey
the obligatory equation (12). 
However, if this relativistic particle is a bit of the Polyakov string,
it should obey another obligatory equation (see Eq.(14)).
 
According to the matrix (8), the covariance of the equation
(12) implies ${\dot \sigma} \neq 0$.
Thus, for a given spacetime and worldsheet, $i.e.$ known $G_{\mu\nu}(X)$
and $X^\mu (\sigma , \tau)$, the equation (12)
gives ${\dot \sigma}$ in terms of $\sigma$ and $\tau$. That is,
the motion of a particle inside its string is influenced by the 
background metric and the geometrical shape of the string. 
In other words, the particle feels an acting potential through a complex
combination of the background geometry and the string shape.
In the next section the equation (12) will be analyzed.

\subsection{The Polyakov string}

The reparametrization invariance can be used to make
the gauge choice $h(\tau) = l^2$, where $l$ is a characteristic
length of the theory. With this gauge the mass term of (7) becomes
constant, and hence we ignore it. Again this system may be any 
one-dimensional physical object. Requesting the Lagrangian of (7) 
be the Lagrangian density of the Polyakov string leads to the
special behavior of the relativistic particles which is described 
by the equation
\bea
H^{ab}h_{ab}=-\frac{l^2}{2\pi \alpha'}\sqrt{g}g^{ab}h_{ab},
\eea
where $g \equiv -\det g_{ab}$. 
This equation is analogue of the equation
(12). According to this, the action (7) gives the following
Lagrangian for a Polyakov particle in the curve $C$,
\bea
{\cal{L}}[X(\sigma_p ,
\tau) , g_{ab}(\sigma_p , \tau)] = - \frac{1}{4\pi \alpha'}
\sqrt{g}g^{ab}(\sigma_p , \tau)G_{\mu\nu}(X)\partial_a X^\mu(\sigma_p ,
\tau)
\partial_b X^\nu(\sigma_p , \tau) .
\eea

Consider an interval $(-\frac{\Delta \sigma_p}{2} , \frac{\Delta
\sigma_p}{2})$ of the string which its center is located at
$\sigma_p$. Thus, its Lagrangian is proportional to
the factor $\Delta \sigma_p$. The total Lagrangian of the string is
given by the summation of the Lagrangians of the intervals, $i.e.$
$\sum_p {\cal{L}}[X(\sigma_p , \tau), g_{ab}(\sigma_p , \tau)]
\Delta \sigma_p$. 
When the infinitesimal length $\Delta \sigma_p$ goes to zero, 
the summation changes to the integral, and hence we obtain the Polyakov
action for the string
\bea
S_P = - \frac{1}{4\pi \alpha'} \int
d\sigma \int d\tau \sqrt{g}g^{ab}G_{\mu\nu}(X)\partial_a X^\mu
\partial_b X^\nu .
\eea

Not that the Polyakov action of the string gives the equation 
\bea
h_{ab}=\frac{1}{2}g_{ab} (g^{cd}h_{cd}).
\eea
This is the equation of motion of the metric $g_{ab}$. Contracting
it by $H^{ab}$ and using the equation (14) we obtain
\bea
H^{ab}g_{ab}=-\frac{l^2}{\pi \alpha'}\sqrt{g}.
\eea
This equation is equivalent to (14), which will be used.

\subsection{The resulted structure for the bosonic string}

The motion of a particle relative to the string curve, is given by 
$V^\mu={\dot
\sigma}\partial_\sigma X^\mu$. Since the variable ${\dot \sigma}$
appears in our calculations, we call ${\dot \sigma}$ as $speed$.
For the rest particle in the string curve, $i.e.$ ${\dot
\sigma}=0$, the only nonzero element of the matrix $H^{ab}$ is
$H^{00}=1$. Thus, having covariance, $i.e.$ appearance of 
the indices ``$a$'' and ``$b$'' in the
actions (6) and (7) and hence in the string actions (13) and
(16), implies the motion of the particle along the string curve. Therefore,
we obtain the following statement: ``{\it If the fundamental string is a
collection of infinite relativistic particles, they should move along the 
string curve. Motion of each of them depends on the background metric 
and geometrical shape of the string. In addition, we shall see that the
distribution of the particles along the string is not uniform.}'' 
\section{Analysis of the particle behavior in the string}

For simplicity we refer to ${\dot \sigma}$ as the particle speed.
In fact, the string bits form a chain of point-like constituents
in which, from the point of view of the observer on the string, they
enjoy a Galilei invariant dynamics. Thus, we can use 
${\dot \sigma}=d\sigma/d\tau$ as speed.

\subsection{Upper and lower bounds of the particle speed}

Before obtaining the explicit form of the speed ${\dot
\sigma}_p$, it is useful to find upper and lower bounds of it.
For simplicity drop the index ``$p$'' from the particle
coordinate $\sigma_p$. For a Nambu-Goto particle the action (6)
implies the inequality $H^{ab}h_{ab} < 0$, which is
\bea
h_{11} {\dot \sigma}^2 +2 h_{01}{\dot \sigma} +h_{00} <0 .
\eea
This inequality also holds for the Polyakov particles, 
which can be seen from the equation (9).

For discussing about the inequality (19) we separate the cases
$h_{11}>0$ and $h_{11}<0$. For those areas of the string
worldsheet with $h_{11}>0$, the range of the particle speed is
given by
\bea
v_- <{\dot \sigma}<v_+ ,
\eea
where $v_+$ and $v_-$ are as in the following
\bea
v_{\pm} = \frac{-h_{01} \pm
\sqrt{-\det h_{ab}}}{h_{11}} .
\eea
In the case in which we receive the simple Lorentz gauge, $i.e.$, the
flat induced metric $h_{00}=-1,\; h_{01}=0$ and $h_{11}=1$, we have
$v_\pm =\pm 1$ and hence $-1 < {\dot \sigma} <1$.

For those areas of the worldsheet which have $h_{11} <0$, when at
least one of the conditions
\bea
{\dot \sigma} > v'_+ ,
\eea
\bea
{\dot \sigma} < v'_- ,
\eea
holds, the inequality (19) is satisfied. 
The speeds $v'_{\pm}$ are defined by
\bea
v'_{\pm} = \frac{h_{01} \pm \sqrt{-\det h_{ab}}}{|h_{11}|} .
\eea

The actual particle
motion, relative to the string, is given by the spacetime vector
$V^\mu={\dot \sigma} \partial_\sigma X^\mu$. Since $\partial_\sigma X^\mu$ 
for both closed and open strings is a known function of $\sigma$ and 
$\tau$, it is sufficient to
obtain explicit form of the speed ${\dot \sigma}$.

\subsection{Speed of a particle of the Nambu-Goto string}

The particles of the Nambu-Goto string have the speed equation
(12). This equation gives the following speeds
\bea
{\dot \sigma}_\pm = \frac{1}{h_{11}} \bigg{[} -h_{01} \pm \sqrt{\bigg{(}
\frac{h_{11}}{(2\pi \alpha' \rho)^2} -1 \bigg{)} \det h_{ab}}\;
\bigg{]} .
\eea
We observe that the background metric, string shape 
and the mass density of the string determine the speeds of the particles.

Since there is $\det h_{ab} <0$ (which can be seen from (13)), 
the square root leads to the condition $h_{11} \leq (2\pi
\alpha' \rho)^2$. For negative values of $h_{11}$ this inequality always
holds. For positive $h_{11}$ we have
\bea
\rho (\sigma, \tau)\geq \frac{1}{2\pi \alpha'} \sqrt{h_{11}
(\sigma, \tau)} .
\eea
We shall observe that, due to the quantum effects, only the equality holds.

\subsection{Speed of a particle of the Polyakov string}

Each of the equations (14) or (18) can be used for investigating the
particle speed of the Polyakov string. In fact, through the equation
(17) they are equivalent. For analogy with the Nambu-Goto string,
we consider the equation (14).
This equation leads to the following 
speeds for the particles of the Polyakov string
\bea
{\dot \sigma}_\pm = \frac{1}{h_{11}} \bigg{[} -h_{01} \pm \bigg{(}
-\det h_{ab} -\frac{l^2 h_{11}}{2\pi \alpha'}\sqrt{g} g^{ab} h_{ab}
\bigg{)}^{1/2} \bigg{]} .
\eea

Compare the speeds (25) and (27). We note that the behavior of the
particles of the Nambu-Goto string is different from the behavior of
the particles of the Polyakov string. There are two
possible speeds for the particles. Therefore, there are 
two kinds of the particles: particles with the speed ${\dot
\sigma}_+$ and particles with the speed ${\dot \sigma}_-$.
Since the direction of the motion of the particles along the string 
should be the same, each string contains one kind of these particles.
Thus, there are two kinds of the Nambu-Goto string and two kinds
of the Polyakov string.

The condition (19) and the equation (14) imply that
$g^{ab}h_{ab}$ is positive. Also we have $\det h_{ab} < 0$.
According to these, the expression under the square root of (27)
for $h_{11} \leq 0$ always is positive. For the case $h_{11} > 0$
we obtain the condition
\bea
\frac{l^2}{2\pi \alpha'} \leq
\frac{-\det h_{ab}}{h_{11}(\sqrt{g}g^{cd} h_{cd})} .
\eea
This inequality
should hold for any $\sigma$ and $\tau$. This implies that the
quantity $\frac{l^2}{2\pi \alpha'}$ is equal or less than the
minimum value of the right-hand-side of (28), $i.e.$,
\bea
\frac{l^2}{2\pi \alpha'} \leq \bigg{(} \frac{-\det
h_{ab}}{h_{11}(\sqrt{g}g^{cd} h_{cd})}\bigg{)}_{min} .
\eea
This determines the upper bound of the length scale ``$l$''.

For both open string and closed string we know the explicit form 
of the function $X^\mu(\sigma , \tau)$.
From the gauge choice of $g_{ab}$ we can also use $g_{ab} =
\eta_{ab}$. Thus, for a given background geometry, $i.e.$ known 
$G_{\mu\nu}(X)$, the right-hand-sides of the equations (25) and
(27) are known functions of $\sigma$ and $\tau$. In other words,
(25) and (27) define differential equations for the function
$\sigma(\tau)$.
\section{Some quantum considerations}

\subsection{The Nambu-Goto case}

For the given coordinates $(\sigma , \tau)$, the action (10) defines the
Lagrangian of a particle inside the string curve. This Lagrangian defines
the following canonical momentum, which is 
conjugate to the particle coordinate $X^\mu (\sigma , \tau)$, 
\bea
\Pi^{(1)}_\mu(\sigma , \tau)= \frac{\rho}{\sqrt{-H^{ab}h_{ab}}}G_{\mu\nu}
(\partial_\tau X^\nu + {\dot \sigma} \partial_\sigma X^\nu).
\eea 
Therefore, equal time canonical quantization gives 
\bea
[X^\mu (\sigma , \tau), \Pi^\nu_{(1)}(\sigma' , \tau)]=i\eta^{\mu\nu}
\delta(\sigma-\sigma').
\eea
In the same way, the action (13) also defines the canonical
momentum
\bea
\Pi^{(2)}_\mu(\sigma , \tau)= \frac{1}{2\pi \alpha'\sqrt{-\det h_{ab}}}
G_{\mu\nu}(h_{11} \partial_\tau X^\nu -h_{01} \partial_\sigma X^\nu).
\eea
This is conjugate to the coordinate of a point of the string, $i.e.$
$X^\mu (\sigma , \tau)$. This leads to the following quantization
\bea
[X^\mu (\sigma , \tau), \Pi^\nu_{(2)}(\sigma' , \tau)]=i\eta^{\mu\nu}
\delta(\sigma-\sigma').
\eea

Since both points refer to the same particle position, 
we should have $\Pi_{(1)}^\mu=\Pi_{(2)}^\mu$, and hence
\bea
[h_{11}-(2\pi \alpha' \rho)^2]\partial_\tau X^\mu =
[h_{01}+(2\pi \alpha' \rho)^2{\dot \sigma}]\partial_\sigma X^\mu. 
\eea
This implies the equations
\bea
&~& \rho (\sigma, \tau) = \frac{1}{2\pi \alpha'}\sqrt{h_{11}(\sigma, \tau)},
\nonumber\\
&~& {\dot \sigma}=-\frac{h_{01}}{h_{11}}.
\eea
These are consistent with the equation (25). However, the particle speed,
extracted from the quantum considerations, is simpler than the 
classical case. Furthermore, we observe that the mass density $\rho$ is 
not constant. In other words, on a time slice, 
the matter distribution on the string is 
inhomogeneous. Note that some string-bit-models reveal 
homogeneouty and some others
give inhomogeneouty for distribution of bits along the string ($e.g.$ 
see Ref. \cite{8,9,10} and references therein).

\subsection{The Polyakov case}

Again equality of the canonical momenta, extracted from the Lagrangian of
a particle in the string ($i.e.$ from the left-hand-side of (14)) and 
Lagrangian of a point part of the string ($i.e.$ from the 
right-hand-side of (14)) leads to the equations
\bea
&~& \frac{l^2}{2\pi \alpha'}\sqrt{g} g^{00}=-1,
\nonumber\\
&~& {\dot \sigma}=\frac{g^{01}}{g^{00}}.
\eea
Combining this speed with one of the speed equations, $e.g.$ (18), 
implies the condition
\bea
g_{00}+2g_{01} \frac{g^{01}}{g^{00}}+g_{11}\bigg{(}\frac{g^{01}}{g^{00}}
\bigg{)}^2 + \frac{l^2}{\pi\alpha'} \sqrt{g}=0.
\eea
That is, equality of the particle speed from the classical and 
quantum point of views gives the condition (37). 
\section{Conclusions}

We assumed a bit-structure for the bosonic string. We considered
each bit as a relativistic point particle.
We obtained the Nambu-Goto form and the Polyakov form of action
of a bit which moves along the string. From these particle
actions we acquired the Nambu-Goto string and the Polyakov
string. 

We achieved the following structure for the fundamental string:
``If a string is made of infinite 
relativistic particles, they are not fixed in
the string. They move with non-constant speeds along the string.
The motion of each of them completely depends on the background geometry
and string shape. In addition, distribution of the particles along the 
string is not uniform''.

Some upper and lower bounds determine the range of the particle
speed ${\dot \sigma}$. 
Explicit forms of the speed indicate that the
speed depends on the position of the particle on the string worldsheet.
This dependence comes via the induced metric of the worldsheet.
Therefore, each of the speed equations provides a differential equation
for the function $\sigma(\tau)$.
The reality of the particle speed, inside the Polyakov string, 
determines an
upper bound for the characteristic length scale of the theory.

We observed that the particles of the Nambu-Goto string have non-uniform
distribution in the string. In addition, we saw
that canonical quantization gives another speed for the 
particle in the string. For a Nambu-Goto particle this speed is
consistent with the classical case. For a Polyakov particle the 
consistency of the speeds puts a condition on the elements of the
intrinsic metric of the worldsheet.

\end{document}